\documentclass[amssymb,prb,twocolumn,showpacs]{revtex4}
\usepackage{epsfig}
\usepackage{dcolumn}
\usepackage{amsmath}
\begin{document}
%\documentclass[12 pt,a4paper]{article} %selecciona el tipo de documento
%\usepackage[english]{babel} %selecciona el idioma
%\frenchspacing %trata los espacios despues de los puntos igual que los otros
%\usepackage{epsfig}
%\usepackage{amsmath}
%\usepackage[a4paper,dvips]{geometry}
%\geometry{textwidth=16 cm, textheight=22 cm}
%\begin{document}
\title{\bf\ Synchronized Scattering and Geometric Dephasing in Microwave-Induced Resistance Oscillations.}
\author{Jesús I\~narrea$^{1}$ }
\address{$^1$Escuela Polit\'ecnica
Superior,Universidad Carlos III,Leganes, Madrid, 28911, Spain.}

%%%%%%%%%%%%%%%%%%%%%%%%%%%%%%%%%%%%%%%%%%%%%%%%%%%%%%%%%%%%%%%%%%%%%%%%%%%%%%
%\section{Abstract}
\begin{abstract}
We present a microscopic quantum transport mechanism for microwave-induced resistance oscillations (MIRO) based on microwave-driven coherent states. By evaluating the impurity scattering matrix element between driven states, we show that the instantaneous scattering rate becomes dynamically synchronized with the velocity of the radiation-induced motion. The resulting velocity-dependent scattering provides a direct microscopic origin for the finite time-averaged transport response underlying MIRO. We further introduce a geometric dephasing mechanism arising from the accumulated displacement of the driven coherent state, which limits the oscillation amplitude at high microwave power. This mechanism naturally explains the observed crossover from linear to sublinear power dependence and predicts a magnetic-field-dependent saturation scale governed by the ratio between the driven displacement amplitude and the cyclotron radius. Our results establish a coherent-state picture of MIRO in which transport is controlled by synchronized scattering and geometric decoherence.

\end{abstract}
%%%%%%%%%%%%%%%%%%%%%%%%%%%%%%%%%%%%%%%%%%%%%%%%%%%%%%%%%%%%%%%%%%%%%%%%%%%%%%
\maketitle
%\section{Introduction}
{\it Introduction.}
\begin{figure}
\centering\epsfxsize=3.in \epsfysize=2.5in
\epsffile{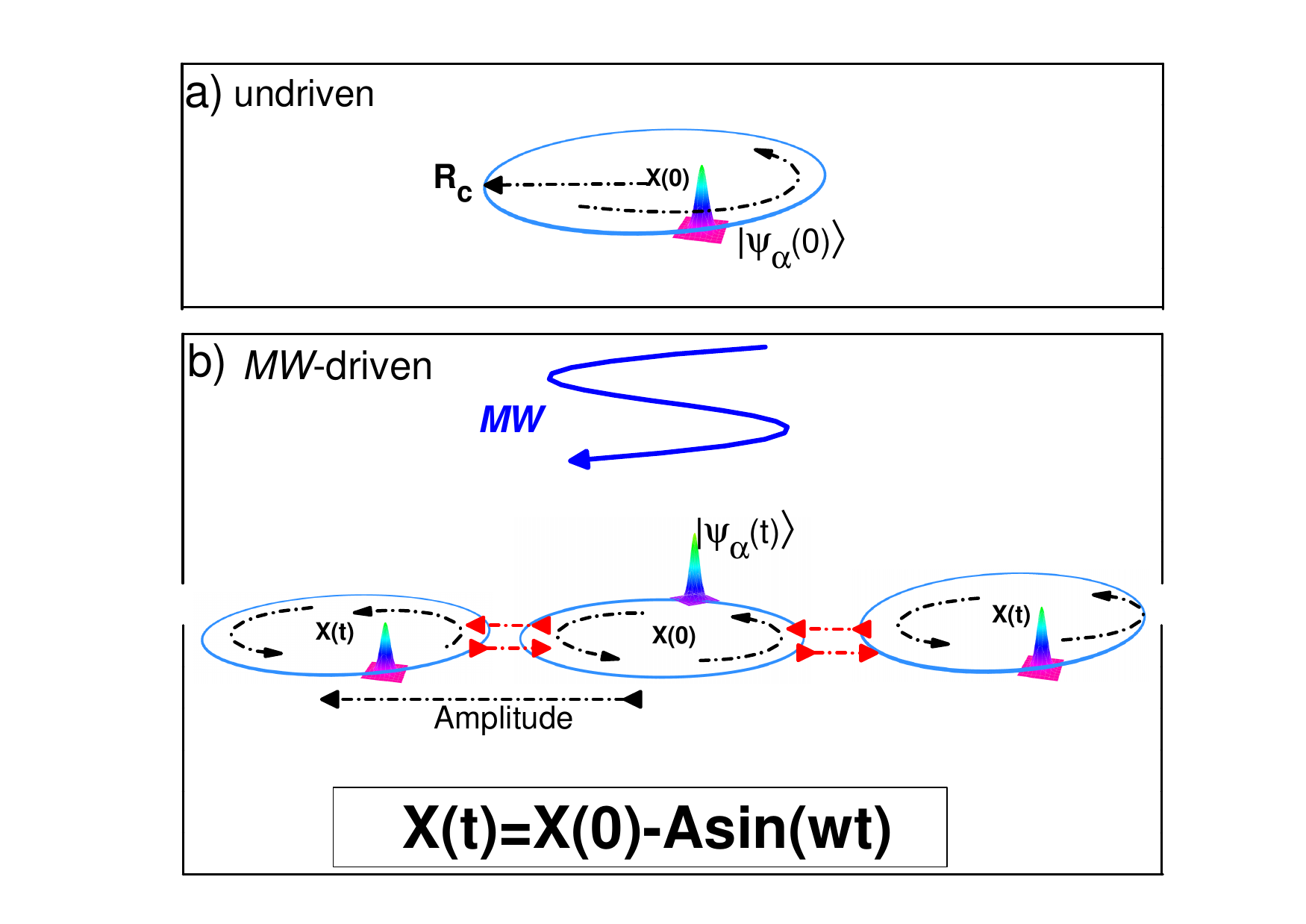}
\caption{Schematic diagrams of the coherent states: the probability density of the coherent state is a
constant-shaped Gaussian
distribution.
%whose center oscillates in a harmonic potential similarly as
%its classical counterpart.
a) The upper  part exhibits the undriven or dark case.
The wave packet displaces harmonically with the cyclotron frequency. b) The lower
part exhibits the driven case, where the wave packet displaces simultaneously with the  cyclotron
and radiation frequencies.}
\end{figure}
The discovery of microwave-induced resistance oscillations (MIRO)\cite{mani,zudov,ryzhii}  and zero-resistance states (ZRS) in high-mobility two-dimensional electron systems (2DES) has triggered a profound re-examination of quantum transport far from equilibrium.
%When subjected to a perpendicular magnetic field ($B$) and continuous microwave radiation ($MW$), the magnetoresistance of the 2DES exhibits macroscopic oscillations governed by the ratio of the radiation frequency to the cyclotron frequency, $\omega/\omega_c$.
%At high radiation intensities, the minima of these oscillations systematically evolve, leading to the formation of states  of zero resistance (ZRS).
To describe the underlying microscopic mechanism, two dominant theoretical frameworks have emerged: the inelastic model\cite{dimi}, which relies on a radiation-induced non-equilibrium distribution function, and the displacement model\cite{girvin}, which focuses on the spatial shift of electrons during impurity-assisted scattering events under the driving microwave field. An
alternative approach is provided by the radiation-driven electron orbit model\cite{ina1,ina2,ina3,ina4}, in which the guiding centers of Landau states-based
coherent states\cite{cohen,inacoher} undergo  harmonic oscillations driven by the microwave field.
Within this picture, the irradiated longitudinal magnetoresistance ($R_{xx}$) comes out from the interplay between this oscillatory
motion and impurity scattering events. Despite its success in capturing key experimental features, a conceptual limitation of this model lies in the implicit assumption of a time-independent scattering rate. In such a scenario, the oscillatory motion of the guiding center averages to zero over time unless phase correlations are effectively preserved during scattering events. This has motivated critical discussions regarding the role of temporal averaging and the persistence of phase-dependent contributions in the transport response.

In this work, based on the coherent states\cite{sro} approach of the radiation-driven electron orbit model\cite{inacoher},  we address this issue by
fully solving the scattering matrix element
with irradiated coherent states. Thus, we obtain a microwave-driven modulation of the scattering probability
which is no longer uniform in time, but instead inherits a periodic component from the radiation field.
The instantaneous scattering rate is fundamentally modulated by the physical velocity of the driven coherent state and
 the interaction with the random impurity potential is maximized at $\omega t = 2n\pi$, with  $n$ being a positive integer.
This corresponds to a situation where the shifting Landau orbits sweep through the crystal lattice at their maximum speed.
Thus, because the scattering probability is synchronized with this velocity maximum, the time-averaged transport equations yield a non-vanishing stationary direct current,
which in the end give rise to MIRO.

% We demonstrate that the total phase accumulated by the shifting Landau orbits during their quantum lifetime $\tau$ can be rigorously decomposed into an integer number $n$ of complete periodic cycles plus a terminal residual phase fraction ($\omega \tau_0$). Since the driving potential preserves time-reversal symmetry over any full period, the intermediate cycles yield zero net current. The spatial symmetry is exclusively broken during the terminal sub-cycle window $\tau_0$ preceding the elastic collapse onto the final Landau state.

% the instantaneous scattering rate is fundamentally modulated by the physical velocity of the driven coherent state ($W_{\text{scat}} \propto \cos\omega t$). The interaction with the random impurity potential is maximized at $\omega t = 2n\pi$, which corresponds precisely to the kinematic peaks where the shifting Landau orbits sweep through the crystal lattice at their maximum speed. Because the scattering probability is synchronized with this velocity maximum, the time-averaged transport equations yield a non-vanishing stationary direct current via an internal quantum rectification process,

We introduce a phenomenological accumulated dephasing mechanism associated with the radiation-driven evolution of coherent states. The dephasing is quantified through the time-integrated overlap between the initial coherent state and the microwave-driven coherent state during the flight time between successive scattering events. Since the latter follows a radiation-driven trajectory, the flight time parameterizes not only the temporal evolution but also the spatial path traveled by the coherent state. Consequently, the accumulated overlap progressively decreases as the state evolves, giving rise to an effective damping of the coherent-state amplitude. This mechanism naturally accounts for the crossover from linear to sublinear power dependence\cite{manipower1,manipower2,jesmanipower,zudovpower}, of the microwave-induced resistance oscillations and predicts a characteristic magnetic-field dependence.

{\it Theoretical model.}
According to the coherent states based extension of the microwave-driven electron orbit model,
 the wave function corresponding to
the coherent state of the MW-driven quantum oscillator is given by \cite{inacoher,inacoher2,kerner,park},
%\begin{equation}
%\psi_{\alpha}(x,t)=e^{\frac{i}{\hbar}\Theta (t)}\langle x| D(\alpha)|\phi_{0}(x-X(0)-x_{o}(t))\rangle
%\end{equation}
%\large
%\begin{widetext}
\begin{eqnarray}
\psi_{\alpha}(x,t) \propto \phi_{0}[x-X(0)-x_{o}(t)-\langle x \rangle(t)]
\end{eqnarray}
where,
%\large
\begin{eqnarray}
\phi_{0}[x-X(0)-x_{o}(t)-\langle x \rangle(t)]=\nonumber\\
\left (\frac{mw_{c}}{\pi\hbar}\right )^{1/4}
e^{-\left[\frac{x-X(0)-x_{o}(t)-\langle x \rangle(t)}{2\Delta x}\right]^{2}}
\end{eqnarray}
%\normalsize
$X(0)$ is the coherent state guiding center,
%\end{widetext}
%\normalsize
$\langle x \rangle(t)$ is the $w_{c}$-dependent position mean value\cite{cohen},
$\langle x \rangle(t)=\sqrt{\frac{2\hbar}{m^{\ast} w_{c}}}|\alpha_{0}|\cos(w_{c}t)$,
$\alpha_{0}$ being the coherent state amplitude\cite{cohen},
$\Delta x$
%=\sqrt{\frac{\hbar}{2m w_{c}}}$
is
the position uncertainty and
\begin{equation}
x_{o}(t)=\left(\frac{e E_{o}}{m^{*}\sqrt{(w_{c}^{2}-w^{2})^{2}+w^{2}\gamma^{2}}}\right)\sin wt
=A\sin wt
\end{equation}
is the solution of the forced and damped classical harmonic oscillator,
where $E_{o}$ is the radiation electric field amplitude, $\gamma$ is a damping factor
for the electronic interaction with acoustic phonons\cite{ina1}.
%${L}$ is the classical Lagrangian,  and $w_{c}$ the
%cyclotron frequency.
Thus,  the wave function turns out to be
the same as a quantum harmonic oscillator ground state where the center is
driven by $x_{0}(t)$ and $\langle x \rangle(t)$\cite{inacoher} .
%Then, the wave packet associated with $\Psi_{\alpha}(x,t)$ is therefore given by:
%\begin{equation}
%$|\psi_{\alpha}(x,t)|^{2}=|\phi_{0}[x-X(0)-x_{o}(t)-\langle x \rangle(t)]|^{2}$.
%\end{equation}
%Accordingly,
%the microscopic physical description of a high-mobility 2DES under low or moderate $B$ would consist of
%constant-shaped Gaussian wave packets harmonically displacing with   $w$ under radiation (see Fig. 1).

To calculate the longitudinal magnetoresistance, $R_{xx}$, we first obtain the longitudinal conductivity ${\sigma_{xx}}$ following
a semiclassical Boltzmann model\cite{ridley,ando,askerov,davies,ihn},\\
% \begin{equation}
$\sigma_{xx}\propto (\Delta X_{MW})^{2}W_{I} $
%\end{equation}
 $W_{I}$ is the electron charged impurity scattering rate.
We consider now that the scattering takes place between irradiated coherent states of quantum harmonic oscillators.
Thus, $\Delta X_{MW}$ is the distance between the guiding centers  of the scattering-involved irradiated coherent states.
According to Fermi's golden rule $W_{I}$ is given by\cite{inacoher},
%\begin{equation}
 $ W_{I}\propto|<\psi_{\alpha^{'}}|V_{s}|\psi_{\alpha}>|^{2}$
%\end{equation}
%where $N_{i}$ is the number of charged impurities, $\psi_{\alpha}$ and $\psi_{\alpha^{'}}$ are the  irradiated
%coherent states  wave functions  corresponding to the initial and final states respectively,
where $V_{s}$ is the scattering potential for charged impurities\cite{ando}.
% $E_{\alpha}$ and $E_{\alpha^{'}}$ stand for the coherent states initial and final energies respectively\cite{cohen}.
The $V_{s}$ matrix element reads \cite{ridley,ando,askerov}:
%\begin{equation}
$|<\psi_{\alpha^{'}}|V_{s}|\psi_{\alpha}>|^{2}=\sum_{q}|V_{q}|^{2}|I_{\alpha,\alpha^{'}}|^{2}$
%\end{equation}
where $V_{q}$ is the Fourier transform of $V_{s}$
and $I_{\alpha,\alpha^{'}}$  the scattering integral  \cite{inacoher,ridley,ando,askerov},
%\begin{widetext}
%\begin{eqnarray}
%I_{\alpha,\alpha^{'}}(t)=
%\int^{\infty}_{-\infty} \psi_{\alpha^{'}}(x-X^{'}(0)-x_{o}(t)-\langle x' \rangle(t')) e^{i q_{x} x}\times\nonumber\\
%\psi_{\alpha}(x-X(0)-x_{o}(t)-\langle x \rangle(t)) dx  \nonumber\\
%\end{eqnarray}
%where the prime indicates final scattering state.
%\end{widetext}
%where,
%\begin{equation}
%\psi_{\alpha}=e^{i\vartheta_{\alpha}}e^{-iw_{c}t/2}e^{\frac{i}{\hbar}\langle p \rangle(t)x} \left (\frac{mw_{c}}{\pi\hbar}\right )^{1/4}
%e^{-\left[\frac{x-X(0)-\langle x \rangle(t)}{2\Delta x}\right]^{2}}
%\end{equation}
%and similar expression for $\psi_{\alpha^{'}}$

After some lengthy algebra, we obtain an expression\cite{inacoher} for $I_{\alpha,\alpha^{'}}$,
%\large
\begin{equation}
  |I_{\alpha,\alpha^{'}}(t)|=e^{-\frac{[\Delta X_{MW}]^{2}}{8(\Delta x)^{2}}}
  e^{-\frac{q_{x}^{2}(t)2(\Delta x)^{2}}{4}}
\end{equation}
%\normalsize
where $q_{x}(t)$is the $x$-component of the electron momentum change

Finally the expression of  $I_{\alpha,\alpha^{'}}$ reads\cite{ridley},
%\large
\begin{eqnarray}
&&|I_{\alpha,\alpha^{'}}(t)|
=e^{-\frac{q^{2}(\Delta x)^{2}}{2}}\times\nonumber\\
&&\left[e^{\frac{ \Delta X_{0}A}{2 l_{B}^{2}} \left( \sin w(t+\tau)-\sin (wt) \right)-\frac{A^{2}}{4 l_{B}^{2}} \left( \sin w(t+\tau)-\sin (wt) \right)^{2}}\right] \nonumber\\
%&=&|I_{\alpha,\alpha^{'}}(0)|e^{-\frac{2 \Delta X_{0}A}{8(\Delta x)^{2}} \left( \sin w(t+\tau)-\sin (wt) \right)}
\end{eqnarray}
where $t$ is the initial  time for the scattering event and $\tau$ is the evolution
time (flight time) between coherent states, i.e., the time it takes to reach
from one state to another\cite{inarashba,inahole,inacoher}.
%\begin{equation}
%|I_{\alpha,\alpha^{'}}|= e^{-\left(\frac{\left(X^{'}(0)-X(0)\right)^{2}}{8(\Delta x)^{2}}+\frac{q_{x}^{2}(\Delta x)^{2} }{2}\right)}
%\end{equation}
%\normalsize
This, in turn, leads us to a final expression for  $W_{I}$,
%\begin{equation}
$W_{I}=W_{I}(0)W_{I}(t)$
%\end{equation}
where
\begin{eqnarray}
W_{I}(t)=\left[e^{\frac{ \Delta X_{0}A}{2 l_{B}^{2}} \left( \sin w(t+\tau)-\sin (wt) \right)}
\right]
\end{eqnarray}
and $W_{I}(0)$ is the stationary (dark) scattering rate\cite{inacoher}.
In this expression the term depending on $A^{2}$ has been neglected because is much smaller than the linear term.
%\begin{equation}
%W_{I}(0)=\frac{n_{i}e^{4}}{2 \pi \hbar \epsilon^{2}}\int \frac{e^{-q^{2}(\Delta x)^{2}}}{(q+q_{TF})^{2}}(1-\cos \theta)\delta(E_{\alpha^{'}}-E_{\alpha})d^{2}q
%\end{equation}
%where $n_{i}$ is the  charged impurity density
%and $\theta$ is the scattering
%angle.
 Therefore, the irradiated coherent state when under scattering gives rise naturally to an
additional time-dependent factor, $W_{I}(t)$, that dramatically  alters the dark scattering rate.

The  magnetic field normally used in MIRO experiments is such that  $\tau/T>1$, $T$, being the radiation period.
This means that during the flight time the irradiated coherent states  performs many MW-driven oscillations,
 in which they  interact with the local environment acting  as a microscopic source of phase friction.
This leads to  dephasing of the driven coherent state.
The accumulated geometric dephasing is quantified through the overlap (fidelity) between the initial coherent state and the same coherent state translated along its microwave-driven trajectory. The progressive reduction of this overlap with the accumulated path length provides the effective dephasing envelope introduced below.
We obtain this overlap
  using the translation operator, $T(s) = \exp[ \frac{s}{R_c} (a^\dagger - a) ]$. The dephasing of the wave packet after performing a total track length $s$ is given by :$\langle \alpha(0) \vert{} T(s) \vert{} \alpha(0) \rangle = \exp\left( -\frac{s^2}{2R_c^2} \right)$. The driven coherent state executes rapid harmonic oscillations under the microwave field, with an instantaneous spatial position $x(t) = A \sin(\omega t)$ and velocity $v(t) =-  A\omega \cos(\omega t)$. The total integrated path length $s$ accumulated by the wave packet during the orbital flight time $\tau$ is determined by the time-integral of its speed:$$s = \int_{0}^{\tau} \vert{}v(t)\vert{} \, dt = \int_{0}^{\tau} \vert{}A\omega \cos(\omega t)\vert{} \, dt \approx A \cdot \omega\tau$$where the last step applies to the multi-oscillation transport regime ($\omega\tau \gg 1$), dropping the rapidly oscillating phase factors.
The overlap between the initial and translated coherent states formally yields a Gaussian attenuation.
Although the coherent-state overlap is Gaussian, over the experimentally relevant parameter range it is accurately reproduced by an exponential envelope. We therefore adopt the latter as an effective phenomenological representation of the accumulated overlap decay.
%However, within the experimentally relevant parameter range, this attenuation is well approximated by a simple exponential decay. We therefore adopt the latter as an effective accumulated dephasing factor, which preserves the essential physical behavior while allowing for a compact analytical description.
Thus, we can write a single space-time dephasing envelope:$F_{\text{env}}=\langle \alpha(0) \vert{} T(s) \vert{} \alpha(0) \rangle \approx \exp\left( -\frac{A  \omega\tau}{R_c} \right)$.
% that we introduce in our model through a temporal dephasing term or decoherence  factor,
%mathematically expressed as $\exp(-\omega\tau)$. Thus, this term accounts for the system's intrinsic loss of quantum coherence through the
%radiation-driven dynamics of the coherent state during $\tau$.

% In the experimental window of low magnetic fields ($R_c \gg 1$) and moderate-to-high powers, this quadratic overlap profile naturally linearizes into a compact, single space-time dephasing envelope:$$F_{\text{env}} = \vert{}\mathcal{A}(s)\vert{}^2 \approx \exp\left( -\frac{A \cdot \omega\tau}{R_c} \right)$$where the coupling constant is absorbed into the effective disorder scale. This unified exponential factor acts as a global multiplier to the transport equations. Under intense driving, where the spatial sweeping amplitude $A$ approaches the cyclotron radius $R_c$, the joint exponent becomes robustly large, forcing $F_{\text{env}}$ to evaluate to a small, well-behaved damping value. This smallness is physically mandatory: it establishes an intrinsic quantum boundary that suppresses the unphysical linear-in-$\tau$ divergence stemming from the standard synchronous scattering sine-difference expansion, $\omega\tau \cos(\omega t)$, thereby naturally dictating the sublinear power crossover observed in experiments without phenomenological parameters

In our model, this dephasing term is affecting the amplitude $A$ and then  the exponent of $W_{I}(t)$ is given by,
\begin{equation}
\frac{ \Delta X_{0}}{2 l_{B}^{2}} Ae^{-\frac{A \cdot \omega\tau}{R_c}}\bigg( \sin w(t+\tau)-\sin (wt) \bigg)
\end{equation}
where we can expand the sine difference in terms of a Taylor series around $wt$,
\begin{eqnarray}
&& e^{-\frac{A  \omega\tau}{R_c}}\bigg( \sin w(t+\tau)-\sin (wt) \bigg )=\nonumber\\
&&e^{-\frac{A  \omega\tau}{R_c}}\bigg ( w\tau\cos wt -
 \frac{w^{2}\tau^{2}}{2}\sin wt -\frac{w^{3}\tau^{3}}{3!}\cos wt+...\nonumber\\
% &&\frac{w^{4}\tau^{4}}{4!}\sin wt+....\bigg )
\end{eqnarray}
The sine difference is expanded around wt. Since the entire modulation is globally weighted by the exponentially decaying dephasing envelope, its overall amplitude rapidly decreases as the accumulated overlap between the initial and driven coherent states decays. Consequently, higher-order corrections to the Taylor expansion contribute only weakly to the transport modulation, so that retaining the leading linear term provides an accurate approximation in the experimentally relevant regime.
%Due to the exponential  suppression, the higher-order terms of the series become physically redundant for large phase arguments. This ensures that the net transport modulation is mainly driven by the small-$\omega\tau$ window, fully justifying a first-order truncation where only the linear term is retained.
%Formally, the total phase accumulated by the driving coherent state  during $\tau$ can be decomposed into $n$ complete periodic cycles
%plus a terminal residual phase: $\omega(t + \tau) = \omega t + 2\pi n + \phi_{\text{residual}}$.
%Since the driving potential preserves spatial and temporal symmetry over any integer number of full periods, the $2\pi n$
%term vanishes identically under the periodic boundary conditions of the sine function.
%The symmetry is broken exclusively by the residual phase accumulated during the terminal fraction of the cycle where scattering occurs,
%defined as $\phi_{\text{residual}} = \omega \tau_0$.
Thus, as a first approximation, the dephasing-weighted sine difference can be reduced to its leading contribution,
\begin{equation}
 e^{-\frac{A  \omega\tau}{R_c}}\bigg (\sin \omega(t+\tau)-\sin (\omega t)\bigg ) \approx e^{-\frac{A  \omega\tau}{R_c}} \omega \tau \cos \omega t
\end{equation}
Substituting this result in $W_{I}(t)$ and Taylor-expanding the
exponential  we get to:
%\begin{eqnarray}
%&&e^{-\frac{ \Delta X_{0}A}{2 l_{B}^{2}} w \tau_{0} \cos wt } e^{-\frac{ A^{2}}{4 l_{B}^{2}} w^{2} \tau_{0}^{2} \cos^{2} wt }\simeq\nonumber \\
%&&\left(1+\frac{ \Delta X_{0}A}{2 l_{B}^{2}} w \tau_{0} \cos wt+..\right)\left(1+\frac{ A^{2}}{4 l_{B}^{2}} w^{2} \tau_{0}^{2} \cos^{2} wt+..\right)\nonumber\\
%\end{eqnarray}
%The scattering rate finally reads like this,
\begin{eqnarray}
W_{I}= W_{I}(0)\times \bigg(1+\frac{ \Delta X_{0}A e^{-\frac{A  \omega\tau}{R_c}}}{2 l_{B}^{2}} w \tau \cos wt +\nonumber\\
\frac{1}{2!}\frac{\Delta X^{2}_{0} A^{2}e^{-\frac{2A  \omega\tau}{R_c}}}{4 l_{B}^{4}} w^{2} \tau^{2}\cos^{2} wt \bigg)
\end{eqnarray}
where we have kept  only the three leading terms.

We can identify the driven coherent state´s velocity ($wA\cos wt$) and this
suggests that the charged impurity scattering is no longer homogeneous during the driven oscillation and takes place
with more intensity where velocity
is maximum. i.e., at the oscillation´s midpoint, $wt=2\pi n$.
This is an important result, because our model reveals a synchronic scattering mechanism, where elastic
collision rates lock dynamically to the driven coherent states.
Physically, the analytical emergence of the $\cos\omega t$ dependence in the scattering rate reflects
the instantaneous velocity of the driving coherent state ($v_{\text{orbit}} \propto \omega \cos\omega t$).
The interaction probability is mathematically maximized at $\omega t = 2n\pi$, corresponding to the exact moments in the microwave cycle
where the coherent state sweeps through space at its maximum speed. At this kinematic peak,
the wave packet collides with the random impurity landscape, maximizing the scattering rate.
This mechanism establishes that the non-zero stationary current is fundamentally rooted in a velocity-dependent quantum scattering.

%We can develop $\delta(E_{\alpha^{'}}-E_{\alpha})$ as a sum of Lorentzian functions and with the use
%of the Poisson sum rules get to a an expression that reads:
%\begin{equation}
%\delta(E_{\alpha^{'}}-E_{\alpha})=\frac{1}{\hbar w_{c}}\left(\frac{1+e^{-\pi\Gamma/\hbar w_{c}}}{1-e^{-\pi\Gamma/\hbar w_{c}}}\right)
%\end{equation}
%\newline

The next step to finally obtain $\sigma_{xx}$  is to calculate  the time average of the product $W_{I}(\Delta X_{MW})^{2}$,
where now $(\Delta X_{MW})^{2}=\left[\Delta X_{0}-A e^{-\frac{A  \omega\tau}{R_c}}(\sin(wt+w\tau)-\sin(wt))\right]^{2}$. 

After some lengthy algebra we obtain
for $\sigma_{xx}$:
\begin{eqnarray}
\sigma_{xx} \propto \bigg\langle W_{I}(\Delta X_{MW})^{2} \bigg\rangle=
%&& \bigg \langle \left[1+\frac{ \Delta X_{0}A}{2 l_{B}^{2}} w \tau \cos wt +\frac{\Delta X^{2}_{0} A^{2}}{4 l_{B}^{4}} w^{2} \tau^{2}\cos^{2} wt\right] \times\nonumber\\
%&&\bigg [\Delta X_{0}-A e^{-\frac{A  \omega\tau}{R_c}}\bigg (\sin(wt+w\tau)-\sin(wt)\bigg ) \bigg ]^{2} \bigg  \rangle \simeq\nonumber\\
\Delta X^{2}_{0}-\frac{2 A_{e}^{2}\Delta X^{2}_{0}}{4 l_{B}^{2}}  w \tau \sin w\tau+\nonumber\\
 \frac{A_{e}^{4} \Delta X^{2}_{0}}{32 l_{B}^{4}} w^{2} \tau^{2}\sin^{2} w\tau \nonumber\\
\end{eqnarray}
where $A_{e}= e^{-\frac{A  \omega\tau}{R_c}}$.
The above result constitutes the core radiation part of $\sigma_{xx}$ and accounts for the rise of MIRO peaks and valleys and zero resistance states.
%Finally, gathering all terms and solving the energy integral,  we obtain an expression for $\sigma_{xx}$ that reads,
%\begin{widetext}
%\begin{eqnarray}
%\sigma_{xx}=\frac{n_{i}e^{6}m^{\ast}}{2\pi^{3}\hbar^{3}\epsilon^{2}}(\Delta X_{0})^{2}\frac{1}{\hbar w_{c}}\left(\frac{1+e^{-\pi\Gamma/\hbar w_{c}}}{1-e^{-\pi\Gamma/\hbar w_{c}}}\right) \times \nonumber\\
%\left(1-\frac{2\chi_{s}}{\sinh (\chi_{s}) } \cos\left(\frac{2\pi E_{F}}{\hbar w_{c}}\right) e^{-\pi\Gamma/\hbar w_{c}} \right)\times\nonumber\\
% \int \frac{e^{-q^{2}(\Delta x)^{2}}}{(q+q_{TF})^{2}}(1-\cos \theta) d^{2}q
%\end{eqnarray}
%\end{widetext}
%where $\chi_{s}=2\pi^{2}k_{B}T/\hbar w_{c}$, $k_{B}$ being the Boltzmann constant, $E_{F}$ the Fermi energy
%and $\Gamma$ the Landau level width.
To obtain
$R_{xx}$ we use the relation
$R_{xx}=\frac{\sigma_{xx}}{\sigma_{xx}^{2}+\sigma_{xy}^{2}}
\simeq\frac{\sigma_{xx}}{\sigma_{xy}^{2}}$, where
$\sigma_{xy}\simeq\frac{n_{e}e}{B}$ and
$\sigma_{xx}\ll\sigma_{xy}$, $n_{e}$ being the 2D electron density.
Consequently, the functional
dependence of the photo-induced magnetoresistance   $ R_{xx}$ on the
MW driving amplitude $A$ is given by,
%\begin{equation}
$ R_{xx} \propto A^{2}e^{-\frac{2 A  \omega\tau}{R_c}}\propto P e^{-\frac{\sqrt{P}  \omega\tau}{R_c}}$.
%\end{equation}
%In the $\sigma_{xx}$ expression we observe that it depends on $E^{2}_{0}$ through the sine term
%which is the most important.  This suggests
 According  to this result we obtain a linear $ R_{xx}$ dependence on $P$
 at low $P$ values and in agreement with   experimental results\cite{manipower1,manipower2,jesmanipower,zudovpower}.
 However, as $P$  rises the $R_{xx}$ curve increasingly bends following a
sublinear law. Above, we explained that the driven coherent state experiences dephasing along
the harmonic displacement. Thus, the longer the displacement or higher $P$, the more intense the coherent state dephasing.
%To account for this effect we introduce a geometric dephasing exponential given by $e^{-A/R_{c}}$.
%The geometric dephasing introduced herein  represents an intrinsic amplitude attenuation of the driven quantum state itself.
%When the coherent state  of the Landau oscillator is driven by the microwave field, it is forced to perform a macroscopic spatial trajectory with displacement amplitude $A$.
As this wave packet sweeps through the long-range impurity landscape, the spatial accumulation of local potential fluctuations alters its internal phase structure,
 leading to a loss of quantum overlap with the unperturbed state.
 %Because the spatial extension of the Landau wave function is physically bounded by the cyclotron radius $R_c$,
% the  amplitude of the driven state decays exponentially as $\exp(-A/R_c)$ when the sweeping distance explores uncorrelated regions of the background potential. Consequently, this geometric envelope directly
%  lessen the macroscopic transport response.
 %  as the net photo-induced current relies entirely on the phase-coherence of these driven states during their dynamic evolution.
%Thus, total
%dephasing consists of a time-dependent part and geometric one.
%Using the dominant power dependence $Rxx∝A2exp(−2Aωτ/Rc$
%), the crossover occurs when A becomes of the order of $0.3Rc/ωτ$.
%The full numerical evaluation of Eq. (11) gives A≃0.23Rc/ωτ."

We can calculate the crossover from the low-power linear scaling to sublinear or the inflection point where
 the rate of growth changes. To determine this crossover threshold
we proceed, as usual, evaluating the second derivative and setting it equal to zero starting from 
the dominant power dependence (see above).
After a straightforward calculation we obtain that the crossover fulfills, $A\simeq 0.23 R_{c}/w \tau$.
Accordingly, at low $B$, $R_{c}$ is large and the crossover displaces at increasing $A$ or $P$. With large
values of $B$ the crossover falls at low $P$ values. In the former case, the curve maintains a linear behavior over a wider range of $P$.
 In the latter, the curve hardly follows a straight line and soon it bends
and becomes sublinear.

\begin{figure}
\centering \epsfxsize=3.4in \epsfysize=3.in
\epsffile{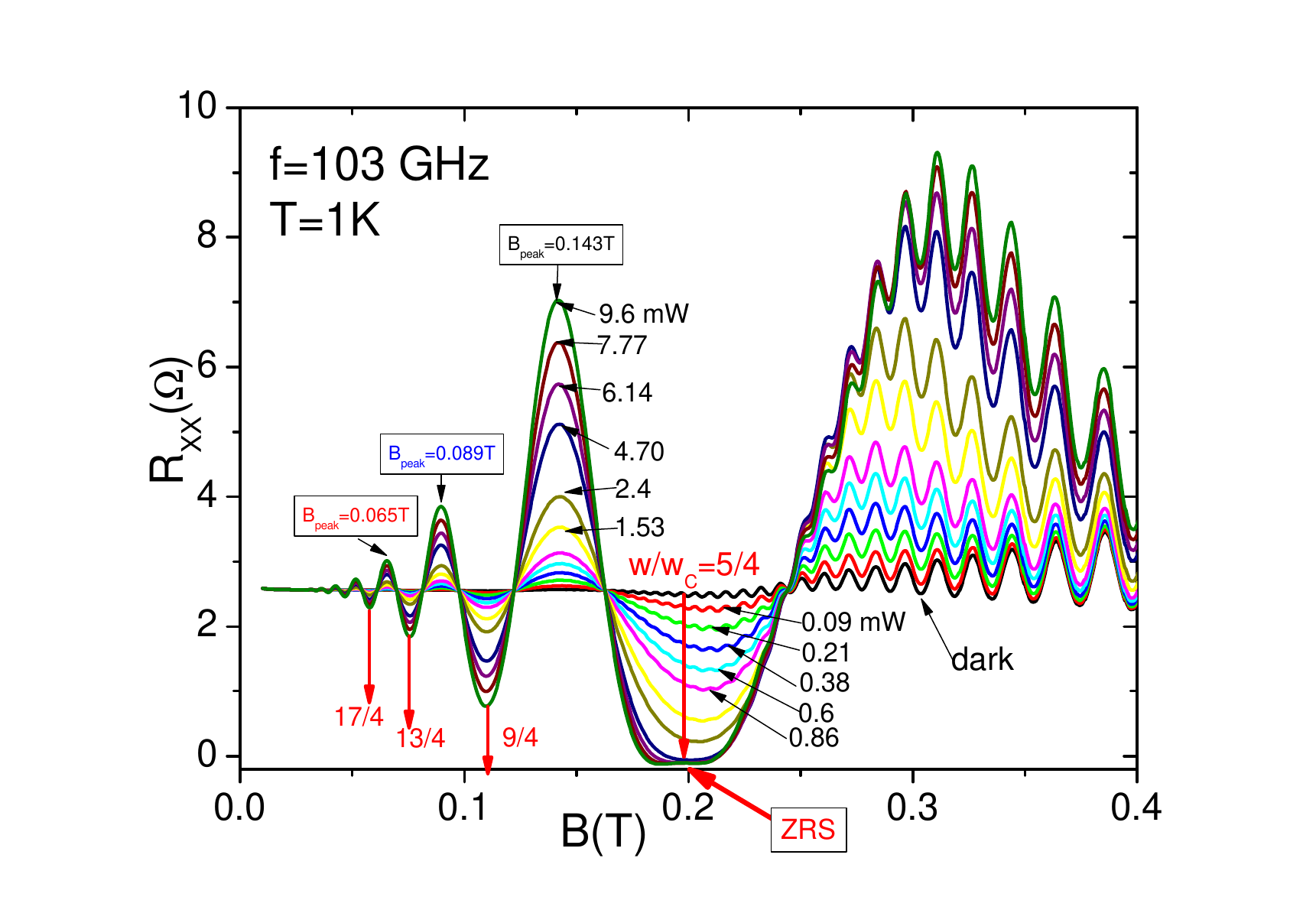}
\caption{Calculated  magnetoresistance   as a function
of $B$,  for a radiation frequency of $103$ GHz and $T=1$ K.
The dark case is also exhibited. Minima positions are indicated with arrows
corresponding to,
$\frac{w}{w_{c}}=j+\frac{1}{4}$, $j$ being a positive integer.
Zero resistance states are obtained around $B\simeq 0.2T$.
As the microwave power increases, the amplitude
of the spatial displacement grows and ends up being reflected in the peak and valley amplitudes. }
%For the latter, increasing power collapses with the rise of ZRS. }
\end{figure}

\begin{figure}
\centering \epsfxsize=3.4in \epsfysize=3.in
\epsffile{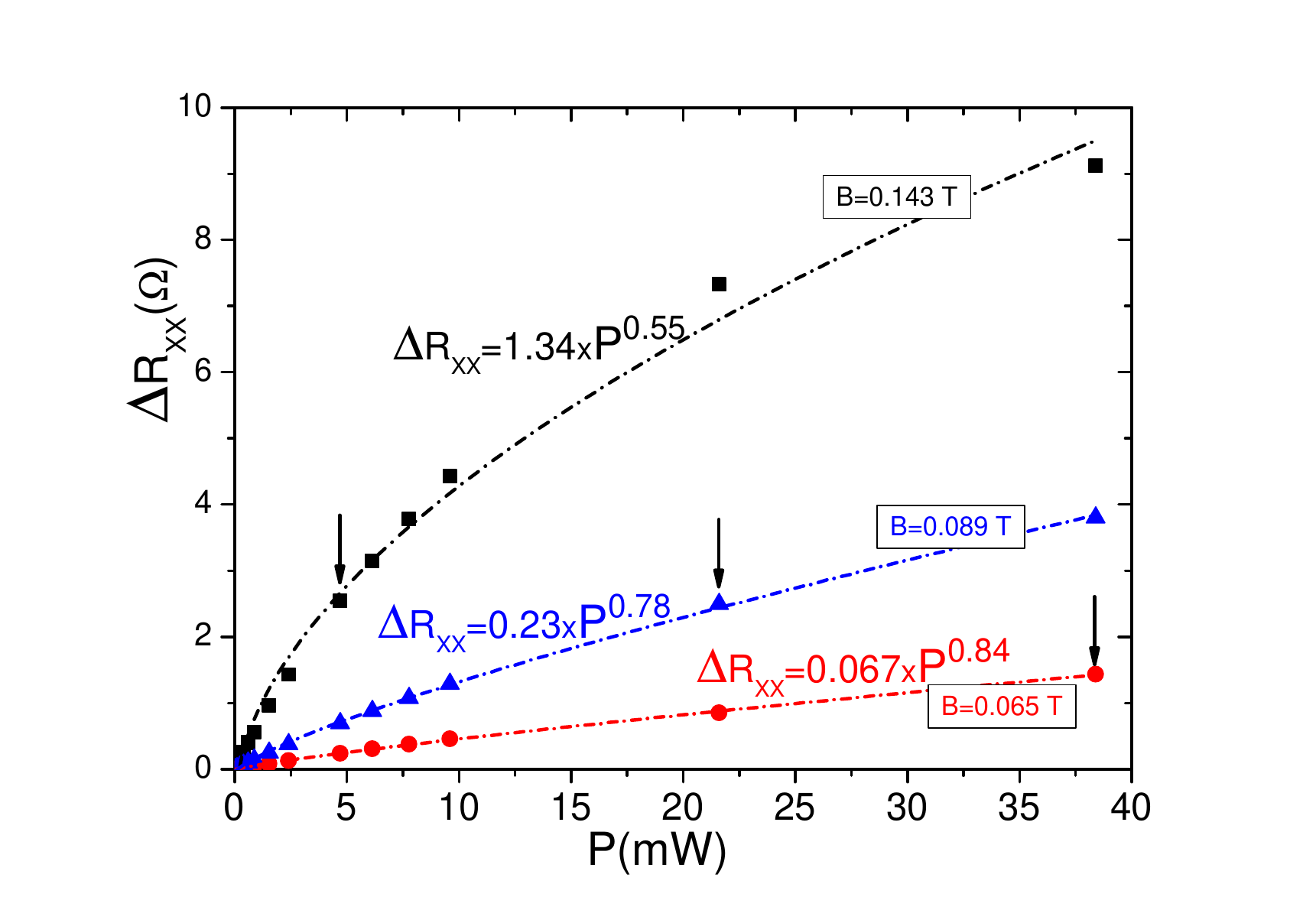}
%\centering \epsfxsize=3.6in \epsfysize=2.in
%\epsffile{newmiro61.pdf}
%\epsfig{file=conducfig1.eps,angle=0,width=0.4\textwidth}
\caption{Calculated amplitude $\Delta R_{xx}=R_{xx}(MW) - R_{xx}(dark)$ of the main $R_{xx}$ peaks of Fig. 2 versus $P$
for $f = 103 GHz$ and  $T=1K$.  Also shown are the fits performed for each MIRO peak. Note that the dependence between $\Delta R_{xx}$ and $P$ is linear first
and starts to bend after the crossover point indicated by an arrow. The arrow is displaced at higher $P$ as
the peak $B$-value decreases. Observe that for increasing $B$, the exponent of the fit increases.
In order to achieve a better fit we have included two extra points of higher $P$, 21.6 and 38.4 mW, not shown in Fig. 2.}
\end{figure}

{\it Results}.
We first focus on the evolution of the longitudinal magnetoresistance as a function of the perpendicular magnetic field  under continuous microwave illumination.
Fig. 2 illustrates irradiated and dark $R_{xx}$ versus $B$ curves for a fixed radiation frequency of $103$ GHz and temperature $T=1$K at several increasing microwave power intensities.
$P$ ranges from $0.09$ to  $9.6$ mW.
As observed, the system exhibits robust resistance oscillations governed by the frequency ratio $w/w_{c}$. Our model accurately reproduces the exact positions of the resonant peaks and valleys according
to experiments\cite{mani,zudov}. Thus, minima positions are indicated with arrows
corresponding to,
$\frac{w}{w_{c}}=j+\frac{1}{4}$, with $j$ being a positive integer.
Zero resistance states are obtained around $B\simeq 0.2T$. This remarkable agreement stems directly from our calculation of the microscopic scattering rate using the driven coherent-state basis.
Thus, we consider that the synchronicity between scattering rate and spatial oscillations of the driven coherent states is at the heart of MIRO.

%As  $P$ increases, the amplitude of the driven-spatial displacement  grows and ends up
%being reflected in the MIRO peak and valley amplitudes. For the latter, increasing power causes MIRO valleys to collapse with
%the rise of ZRS. Higher microwave power corresponds to a larger radiation electric field,
%which translates into a higher coherent state displacement and a higher sweeping velocity of the wave packet through the crystal lattice.
%As explained above, the scattering probability is synchronized with the oscillations of the MW-driven orbits and when $wt = 2n \pi$ the orbits sweep at maximum velocity.
%the quantum rectification process is amplified.
%This explains the initial, pronounced enhancement of the MIRO amplitude observed in Fig. 1 as the system is driven further away from thermal equilibrium.

%B. Power Crossover and the Space-Time Duality of Dephasing
%To unravel the microscopic constraints limiting this radiation-driven
%back and forth motion, we systematically analyze the non-linear power dependence of the MIRO peaks amplitude.
%Thus,
Fig. 3 depicts the photo-induced magnetoresistance amplitude, $\Delta R_{xx}=R_{xx}(MW) - R_{xx}(dark)$,
plotted as a function of the microwave power for three distinct magnetic field values: B = 0.143 T, B = 0.089 T, and B = 0.065 T which correspond to the
positions of the  three main peaks in Fig 2.
The  curves in Fig. 3 reveal a universal trend: at low microwave intensities, $\Delta R_{xx}$ scales strictly linearly with power, ($\Delta R_{xx} \propto P$).
However, as $P$ continues to grow, the system undergoes a crossover, entering a sublinear saturation regime.
%Crucially, this crossover is highly sensitive to the magnetic field landscape, occurring at significantly lower power thresholds as B decreases.
%Our dual dephasing term provides a clear, parameter-free explanation for this space-time crossover.
%While temporal phase friction $(exp^(-w\tau))$ slows the overall amplitude due to high-frequency oscillation during the electron's flight,
%it remains independent of the driving amplitude. The non-linear saturation is entirely driven by the spatial component: the geometric dephasing term, $exp^(-A/Rc)$.
%At low fields ($B = 0.065$ T),  the cyclotron radius of the coherent state is large and at the same time
%the peak amplitude is small (see Fig. 3). Then  geometric dephasing term, $e^{-(\frac{A}{R_{c}})}$,
%remains small in the whole range of radiation power used.
%Remember that $A$ increases as the resonance condition $W^{2}_{c}-w^{2}$ is
%getting closer and this happens at larger $B$.
%Then, the irradiated magnetoresistance remains linear with $P$ almost irrespective of the $P$ intensity.
%Thus, the crossover is obtained at a higher $P$ (see Fig. 4).
%As $B$ increases, ($B = 0.089$ T) the corresponding peak is more intense, ($A$ increases) but the cyclotron radius decreases.
% and the geometrical dephasing term becomes  smaller.
%Thus, according to the above formula, ($A=0.23 R_{c}$),
% the $\Delta R_{xx}$ curves begin to bend and reach the crossover at smaller $P$. For instance,
%for $B=0.143 T$, $\Delta R_{xx}$ stops being linear with $P$ very soon.
Thus, the interplay between $B$ and $A$ is crucial for the  coherent state dephasing.
The fact that our analytical model reproduces the experimental bending\cite{manipower1,manipower2,jesmanipower,zudovpower} indicates that MIRO saturation
 can arise from a coherent geometric mechanism without invoking electron heating.
%When A ~ Rc, the shifting wave packet experiences severe spatial phase averaging as it sweeps across the random impurity landscape. This geometric distortion squashes the coherence of the quantum states. Therefore, at B = 0.0655 T, the geometric damping factor exp(-A/Rc) activates much earlier, forcing the linear-to-sublinear power crossover to take place at a fraction of the intensity required for higher fields (such as B = 0.143 T, where lb = 67.85 nm and the orbits are more compact).
To date, while the power dependence of MIRO has been studied at specific, isolated magnetic fields\cite{manipower1,manipower2,jesmanipower,zudovpower},
a systematic experimental mapping of the linear-to-sublinear crossover as a function of the magnetic field remains absent in the literature.
Therefore, our dephasing framework offers a testable prediction for future magnetotransport experiments.

{\it Summary}.
In summary, we have developed a rigorous non-equilibrium quantum transport framework for microwave-induced resistance oscillations (MIRO)
based on the coherent states extension of the microwave-driven electron orbit model.
We have demonstrated that the irradiated microscopic scattering rate is fundamentally determined by the instantaneous sweeping velocity of the
MW-driven coherent state. This synchronization links the transport response to the kinematic maxima of the wave packet motion, resulting in a finite stationary current.
Furthermore, our model introduces a dual dephasing architecture that establishes a space-time duality in quantum decoherence.
While high-frequency oscillations account for temporal phase dephasing, the non-linear amplitude saturation is governed by a spatial geometric dephasing term.
%This term naturally dictates the linear-to-sublinear power crossover as the displacement amplitude $A$ approaches the cyclotron radius.
% We have shown that this geometric constraint becomes  dominant at larger magnetic fields.

%Finally, our parameter-free agreement with experimental data across different magnetic fields demonstrates that MIRO saturation is a purely coherent, geometry-driven phenomenon rather than a thermal artifact. Beyond the physics of two-dimensional electron systems, the mechanism of velocity-dependent scattering uncovered in this work aligns with universal principles of Floquet engineering and dynamic dissipation found in optical lattices and Rydberg atoms. This positions radiation-driven magnetotransport as an ideal and exceptionally clean solid-state laboratory for exploring universal quantum rectification and non-equilibrium coherent phases.

%In turn, the latter states can become Schr\"odinger cat states when $\alpha$ is large.

%In the same way, we can think of cat states with three o more subcomponents

%and the existence of zero resistance states have been explained.

%We acknowledge useful discussions with
% G. Platero.

This work was supported by the MCYT (Spain) grant PID2023-149072NB-I00.

\end{document}